# Predicting Berborite as Potential Deep-Ultraviolet Nonlinear Optical Crystal from First Principles

Lei Kang,[1,2] Pifu Gong,[1] Zheshuai Lin,[1]* Bing Huang[2]†

[1] *Technical Institute of Physics and Chemistry, Chinese Academy of Sciences, Beijing 100190, China*
[2] *Beijing Computational Science Research Center, Beijing 100193, China*

E-mails: zslin@mail.ipc.ac.cn (Z.L.); bing.huang@csrc.ac.cn (B.H.)

**Following our *ab initio* nonlinear optical (NLO) materials design guidelines, in this Letter, we discovered a novel type of structure to realize potential deep-ultraviolet (DUV) NLO performance in the classical beryllium borate system. By densely stacking the NLO-active layered frameworks, the key design scheme for the structural evolution from the $(Be_2BO_3F_2)_\infty$ layers in $KBe_2BO_3F_2$ (KBBF) to the novel $(Be_2BO_5H_3)_\infty$ layers in berborite is illustrated. Based on available experimental results and systematical theoretical evaluation from first principles, the NLO properties of berborite are further obtained as comparable as the only pratical DUV NLO crystal KBBF. It is demonstrated that berborite can achieve available DUV phase-matched output with strong NLO effect for the practically important 177.3 nm and 193.7 nm lasers. Once obtained with sizable single crystal, it can be applied as a promising DUV NLO crystal.**

Deep-ultraviolet (DUV, $\lambda \leq 200$ nm) nonlinear optical (NLO) crystals are important optoelectronic materials for providing scientists with powerful harmonic laser sources for industrial production (e.g., 193.7 nm lithography) and fundamental research (e.g., 177.3 nm photoelectron spectroscopy) [1-9]. However, DUV NLO crystals are rare because it is extremely challenging to achieve phase-matching (PM) condition in the DUV region [10-12]. The challenge arises from the stringent DUV NLO performance criteria, including (i) large optical bandgap $E_g \geq 6.3$ eV; (ii) sufficient birefringence $\Delta n \geq 0.08$; (iii) effective second-harmonic generation (SHG) effect $d_{\rm eff} \geq d_{36} = 0.39$ pm/V of the benchmarked $KH_2PO_4$ (KDP); and (iv) suitable crystal growth property [11]. So far, $KBe_2BO_3F_2$ (KBBF) is the only practical DUV NLO material that satisfies these criteria well, although it is experiencing the problem of large-sized crystal growth [13].

In the latest decade, to meet the needs of DUV NLO materials exploration, we have proposed a complete set of structure-property relations and *ab initio* design scheme to understand and predict DUV NLO performance more accurately, efficiently and intuitively [14-17] (for more computational details see Methods in the Supporting Information). Based on this systematic NLO materials design platform, on the one hand, we first proposed an important "NLO-size-effect", [18] and accordingly designed a series of NLO structures with excellent DUV PM capabilities via the inter-layer *A*-site cationic regulation mechanism, including $NH_4Be_2BO_3F_2$ (ABBF) [19], $\gamma$-$Be_2BO_3F$ ($\gamma$-BBF) [19], $Be_2CO_3F_2$ (BCF) [20], etc. In particular, ABBF and $\gamma$-BBF have been synthesized in subsequent experiments and indeed showed the promising DUV NLO properties that we predicted [21]. On the other hand, via the intra-layer substitution of anionic groups, we also proposed a series of fluorocarbonates (e.g., $RbAlCO_3F_2$) as the first system that can break through the performance limit reached by KBBF theoretically [22,23]. Recently, we have provided a DUV NLO materials modelling blueprint, aiming to accelerate the process of DUV NLO materials discovery [17]. With this in hand, we can find a shortcut to new DUV NLO materials by effectively evaluating the DUV NLO performance in a fast way combined with preliminary experiments.

In this Letter, based on this design blueprint, we continue to take a different approach to realize the DUV NLO performance. Our original design is still focused on the conventional beryllium borate NLO system and based on the classical KBBF structure. We propose a key design scheme for the structural evolution from the $(Be_2BO_3F_2)_\infty$ layers in KBBF to the novel $(Be_2BO_5H_3)_\infty$ layers by densely stacking the NLO-active layered frameworks as illustrated in Figure 1. Accordingly, we have indeed discovered an existing compound, namely berborite, with a novel type of structure that can exhibit excellent DUV NLO properties. Detailed analysis of the structure-property correlation and the evaluation for the practical 177.3 nm and 193.7 nm laser output are further performed.

Figure 1 schematically illustrates our design proposal, through a concise form of the layer-structural evolution from $(Be_2BO_3F_2)_\infty$ in KBBF to novel $(Be_2BO_5H_2)_\infty$ and $(Be_2BO_5H_3)_\infty$ with local coordination tetrahedral modification via the anionic substitution and vdW engineering. As the starting point of DUV NLO materials modeling, Figure 1a shows the simplified KBBF structure with $(Be_2BO_3F_2)_\infty$ layers as the main components that can produce the DUV NLO performance, where the polar-aligned $(BO_3)^{3-}$ triangles contribute to the SHG effect ($d_{16} \approx 0.45$ pm/V), the coordinated $(BeO_3F)$ tetrahedra enlarge the borate bandgap ($E_{\rm g} \approx 8.3$ eV) by eliminating the nonbonding states on the dangling O atoms of $(BO_3)^{3-}$, and the quasi-planar $(BeO_3F\text{-}BO_3)_\infty$ skeleton enhances the optical birefringence ($\Delta n \approx 0.088$ at default 400 nm) [13]. For details of the local $(BO_3\text{-}BeO_3F)$ anionic structure see Figure 1d.

The detailed NLO properties and crystallographic data are shown in Table 1 and Table S1 of the Supporting Information, with a good agreement between the calculated and experimental results.

Keeping the DUV NLO-active $(Be_2BO_3F_2)_\infty$ layers unchanged, we can obtain such $(Be_2BO_5H_2)_\infty$ layers by taking the anionic substitution of $(OH)^-$ for $F^-$ while maintaining the $D_3$ symmetry, where the hydrogen is randomly occupied with a probability of 1/3 for each site as illustrated in Figure 1b and 1e. Since $(OH)^-$ and $F^-$ have similar valence electron structures and ionic sizes, in certain case this substitution is rational. However, compared with the Be-F bond, the Be-O bond coupling is relatively weaker and the bond length is slightly longer. Therefore, the bandgap of $(Be_2BO_5H_2)_\infty$ may decrease and the anisotropy becomes lower (*i.e.*, birefringence becomes smaller). This conclusion is consistent with our first-principles calculations for $KBe_2BO_5H_2$, which is designed by using $(OH)^-$ instead of $F^-$ as depicted in Figure S1a of the Supporting Information. Table 1 shows that $KBe_2BO_5H_2$ exhibits close SHG effect ($d_{16} \approx 0.46$ pm/V), but smaller bandgap ($E_g \approx 6.6$ eV) and birefringence ($\Delta n \approx 0.057$) than KBBF. As a result, it cannot achieve the PM output in the DUV region ($\lambda_{PM} \approx 250$ nm) [17].

Further applying the *A*-site cationic regulation strategy based on the "NLO-size-effect", we succeed in obtaining two kinds of new layered structures by eliminating the *A*-site $K^+$ cations as depicted in Figure 1c. One kind is derived from the F-bridge-bonded γ-BBF structure by replacing F with OH, [21] and the other from the van der Walls (vdW) connected BCF structure by removing inter-layer $K^+$ [20] (see Figure S1b and Figure S1c, respectively, in the Supporting Information). These two design strategies are expected to further enhance the DUV NLO capability.

The designed $Be_2BO_4H$ has similar bridge-bonded structure as γ-BBF [21]. Due to the anisotropy of bridged $(OH)^-$ units, the designed structure has a large deviation from the ideal pattern with extremely low symmetry (see Figure S1d in the Supporting Information). It indeed opens up an enlarged bandgap ($E_g \approx 7.6$ eV) and birefringence ($\Delta n \approx 0.085$) compared to $KBe_2BO_5H_2$ as listed in Table 1. Accordingly, its shortest PM SHG wavelength $\lambda_{PM} \approx 183$ nm, and has been available for the DUV 193.7 nm output with larger SHG effect $d_{16} \approx 0.55$ pm/V than KBBF, demonstrating the effectiveness of this kind of design scheme. A similar case is $Zn_2BO_4H$ with $P2_1$ symmetry, but it is opaque for the DUV light ($\lambda_{UV} \approx 204$ nm) [24]. In fact, $Be_2BO_4H$ has an existing structure phase, a mineral called hambergite [25]. Unfortunately, it is centrosymmetric (symmetry: *Pbca*) without SHG signal, although its birefringence (exp. $\Delta n \approx 0.074$) is large enough and UV absorption edge (cal. $\lambda_{UV} \approx 160$ nm) is below 200 nm.

The second kind of structure is vdW-densely-stacked, with similar layered frameworks as BCF [20]. The possible structures in the inorganic crystal structure database (*e.g.*, ICSD) actually belong to the phases of berborite [26,27], a beryllium borate mineral with the chemical formula $Be_2BO_5H_3$ (named BBH). Berborite has a good crystal transparency and large uniaxial birefringence (exp. $\Delta n \approx 0.095$), which usually occurs in hexagonal 1T, 2T, 2H polytypes in the nature [27]. The experimental crystal structures exhibit different symmetries of *P3*, *P3c1* and $P6_3$ according to the stacking form and symmetrical operation, as shown in Figure 2a, 2b and 2c.

In the 1T-BBH structure, the covalent layers are layer-by-layer stacking along the *c*-axis with two rotation symmetry (space group *P3*) so there is only one $(Be_2BO_5H_3)_\infty$ monolayer in each unit cell with parallel-aligned polarization in the *a*-*b* plane (see Figure 2a and 2d), which may exhibit potential DUV NLO performance. The first-principles calculations demonstrate that it exhibits very wide bandgap ($E_g \approx 8.2$ eV), strong SHG effect ($d_{16} \approx 0.52$ pm/V) and large birefringence ($\Delta n \approx 0.087$), in good agreement with the available experimental results from Mineral Data Publishing (see Table 1) [28]. Accordingly, it can indeed achieve the DUV SHG output down to 175 nm, which is comparable to the cal. $\lambda_{PM} \approx$ 172 nm of KBBF and thus available for the practical 177.3 nm harmonic laser source [7]. Note that the hydrogen-site is randomly occupied with a probability of 1/2, and some O-H-O bonds between $(Be_2BO_5H_3)_\infty$ layers are formed in addition to the vdW interaction. Therefore, the inter-layer covalent connection of this 1T phase might be enhanced than the inter-layer ionic interaction in KBBF, which would weaken the slippage and is beneficial to the growth along the *c*-axis. Theoretical calculations on elastic properties (see Table S2 in the Supporting Information) show that its bulk modulus is smaller than that of KBBF [29], so its hardness is relatively smaller (exp. ≈ 3 on the Mohs Scale); but its Young Modulus along the *c*-axis is larger than that of KBBF, so its inter-layer mechanical connection becomes to be larger.

It is worth emphasizing that 1T-BBH is a structure with excellent DUV NLO performance comparable to KBBF. In addition to seeing this point from the above discussion including the structure-property evolution as illustrated in Figure 1, we can also confirm it by analyzing the intrinsic NLO origin in details as plotted in Figure 3, especially as compared with KBBF. Figure 3a plots the partial density of states (PDOS) of 1T-BBH, in which the *p* orbitals of O, B and Be are overlapped in the top of valance band (VB) and the bottom of conduction band (CB). Thus, the nonbonding states around the forbidden band edges are saturated to some extent so that bandgap can be opened up and larger than some simple borates. This is very similar to the KBBF case as depicted in Figure 3b. The difference is that the O-H coupling in 1T-BBH further eliminates the dangling bonds on the oxygen atoms, while the fluorine plays the role in KBBF. Based on the electronic structures of the state distribution, the frontier orbitals of 1T-BBH and KBBF are further obtained as illustrated in Figure 3c, from which the Highest Occupied Molecular Orbital (HOMO, orange) and the Lowest Unoccupied Molecular Orbital (LUMO, green) are highlighted to show the dominate origin of the optical transition near the band edge. It can be found that the $(BO_3)^{3-}$ groups mainly contribute the frontier orbitals.

The conclusion can also be illustrated from the analysis of SHG-weighted charge density, which is employed to more intuitively identify the orbitals that contribute to the SHG effect [30-32]. In this methodology, the SHG coefficient is ‘decomposed’ onto the respective orbital or band according to a ‘band-resolved’ scheme, and then the SHG-weighted bands are used to sum the charge densities of all occupied or unoccupied states. As a result, the electronic states irrelevant to SHG are not displayed in the occupied or unoccupied SHG density, and the orbitals vital to SHG are highlighted in the real space. It is clearly illustrated from Figure 3d that the SHG density is mainly located in B and O atoms, indicating that the dominant contribution of the $(BO_3)^{3-}$ to the SHG effect.

To characterize the bond strength between the covalent layers in 1T-BBH and KBBF, we investigated the charge transfer based on the electronic density difference and the inter-layer binding energy. Figure 3e shows that 1T-BBH’s inter-layer charge transfer is more obvious than KBBF, suggesting stronger inter-layer connections. Also, the calculated inter-layer binding energy of 1T-BBH is larger than KBBF. These all indicate more favorable crystal growth performance of 1T-BBH than KBBF especially along the thickness direction.

The in-depth analysis of the points is consistent with our understanding at the beginning. Therefore, 1T-BBH does exhibit excellent DUV NLO performance. As such, next, we are very necessary to evaluate its practical phase-matching capability in the DUV region. We have obtained its Sellmeier dispersion equations by fitting the refractive indices ($n_o$ and $n_e$) and PM angles, which are accurate in the UV and DUV regions from 152 nm to 400 nm as follows:

$$n_o^2 = 2.39925 + \frac{0.00892}{\lambda^2 - 0.01229} - 0.00017\lambda^2$$
$$n_e^2 = 2.15255 + \frac{0.00632}{\lambda^2 - 0.01128} - 0.00012\lambda^2$$

For the important 177.3 nm SHG output, the simulated PM angle $\theta_{PM} \approx 79.4°$ for 1T-BBH, slightly larger than that of KBBF (≈ 77.5°). In the practical type-I PM process (o + o → e), the effective SHG coefficient $d_{\text{eff}}=d_{16}\times\cos\theta_{PM}\times\cos3\psi$, where $\psi$ can be tuned as zero in the actual situation. Accordingly, the $d_{\text{eff}}$ at 177.3 nm of 1T-BBH is comparable to that of KBBF [13]. Similarly, for the important 193.7 nm SHG output, the simulated PM angle $\theta_{PM} \approx 62.1°$ for 1T-BBH, slightly smaller than that of KBBF (≈ 63.1°). As a result, the $d_{\text{eff}}$ at 193.7 nm of 1T-BBH is 1.3 times larger than that of KBBF. In a word, as plotted in Figure 3f, 1T-BBH can realize comparable DUV PM conversion as KBBF under the same application conditions (*e.g.*, laser power and crystal quality).

Different from the 1T phase, 2T-BBH (*P3c1*) and 2H-BBH ($P6_3$) have glide and screw symmetry, respectively, except for the intra-layer rotation. Therefore, each of their unit cells has two monolayers densely stacked and rotated by 60°, as displayed in Figure 2b, 2e and 2c, 2f, which results in the total polarization of bilayers to be offset. Consequently, their SHG effects will become smaller and mainly along the *c*-axis. Our calculations confirm this result that there are only $d_{33}$ equal 0.14 and 0.25 pm/V, respectively. However, their basic building blocks and structural anisotropy have not been changed, so they still have large bandgaps ($E_g \approx 7.7$ eV) and birefringence ($\Delta n \approx 0.09$). Their shortest PM SHG output can reach to 185 and 190 nm, respectively, which can be used for the 193.7 nm output, although their SHG conversion efficiency may be low.

In conclusion, based on our developed *ab initio* modeling blueprint we have predicted berborite as a possible DUV NLO crystal candidate. Detailed first-principles analysis demonstrate that it indeed exhibits a large optical bandgap, strong SHG effect, and sufficient birefringence, so that it can achieve available DUV phase-matched output for the practically important 177.3 nm and 193.7 nm laser sources. Considering its comparable $d_{\text{eff}}$ to KBBF, once obtained with large-sized crystal, it can be applied as a promising DUV NLO crystal with high conversion efficiency.

**Acknowledgements:** We thank H. Zhou for helpful discussion. This work is supported by the NSFC (Grant Nos. 11704023, 51890864, 11634003) and NSAF (Grant No. U1930402) and the Science Challenge Project (Grant No. TZ2016003).

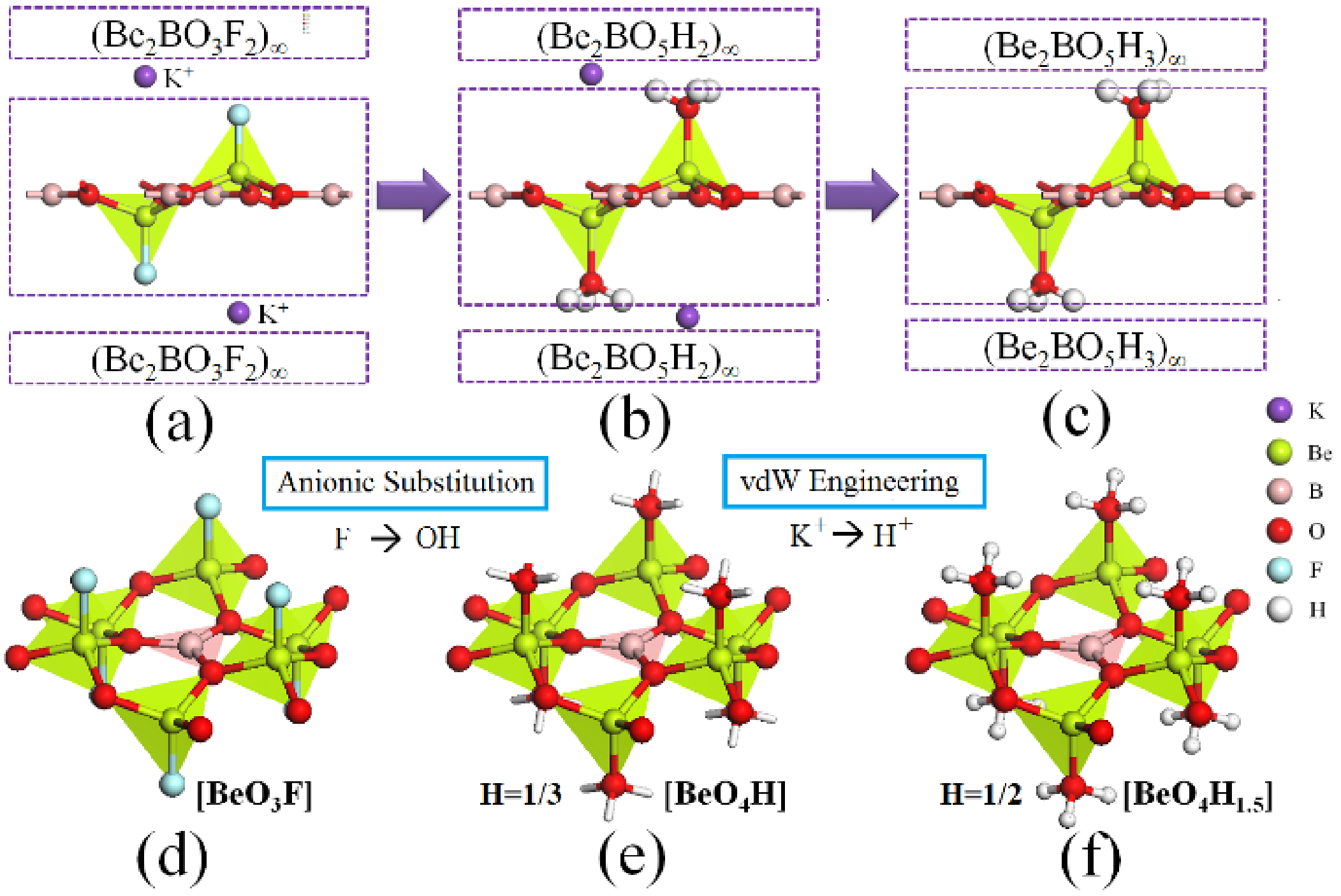

FIG. 1. Layer-structural evolution from $(Be_2BO_3F_2)_\infty$ (a) to $(Be_2BO_5H_2)_\infty$ (b) and $(Be_2BO_5H_3)_\infty$ (c) and local coordination tetrahedra-structural evolution from $(BeO_3F)$ (d) to $(BeO_4H)$ (e) and $(BeO_4H_{1.5})$ (f) via the anionic substitution and van der Waals (vdW) engineering

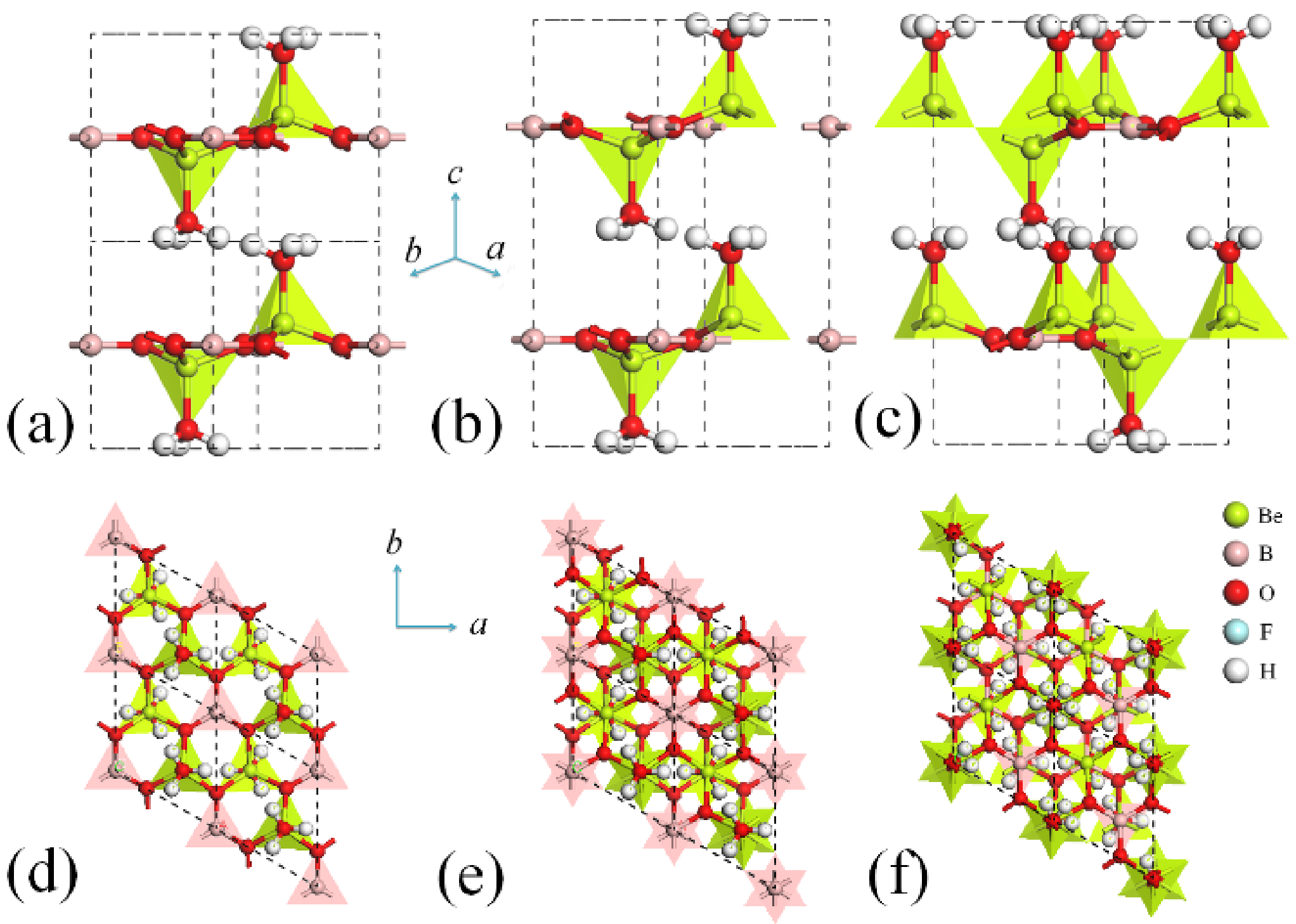

FIG. 2. Experimental crystal structures of berborite phases (*i.e.*, 1T, 2T, 2H) with different symmetries of *P3* (a), *P3c1* (b) and $P6_3$ (c); their layered framework structures within the *a*-*b* plane are displayed in (d), (e) and (f), respectively, from the top view

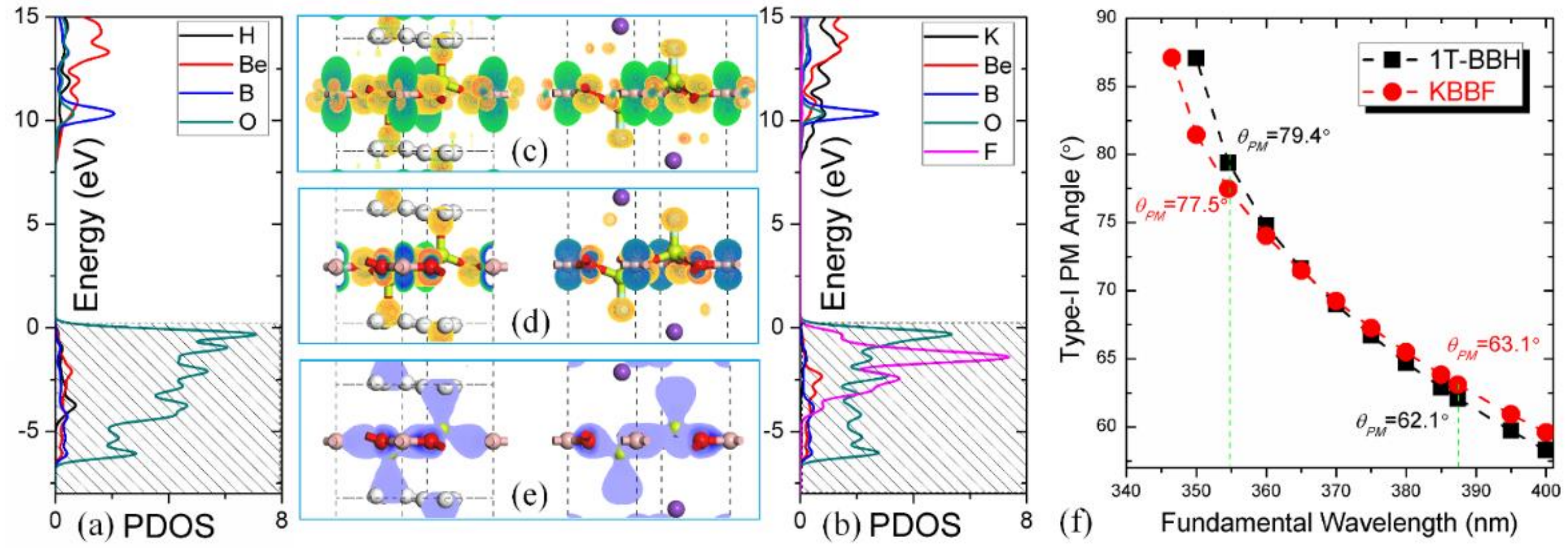

FIG. 3. PDOS in 1T-BBH (a) and KBBF (b), as well as the comparison of frontier orbitals (c), SHG-weight density (d), electronic density difference (e), and type-I PM angles (f) between these two crystals

Table 1. DUV NLO Properties of $KBe_2BO_3X_2$, $Be_2BO_3X$ ($X$ = F, OH) and $Be_2BO_5H_3$ Phases

| | Symmetry | | $E_g$ (eV) | $d_{ij}$ (pm/V) | $n_x$ ($n_o$) | $n_y$ ($n_o$) | $n_z$ ($n_e$) | $\Delta n$ | $\lambda_{UV}$ (nm) | $\lambda_{PM}$ (nm) |
|---|---|---|---|---|---|---|---|---|---|---|
| $KBe_2BO_3F_2$ | Exp. | *R32* | 8.30 | $d_{16}$=0.47 | 1.4920 | 1.4920 | 1.4044 | 0.088 | 147 | 161 |
| | Cal. | *R32* | 8.33 | $d_{16}$=0.42 | 1.4982 | 1.4985 | 1.4315 | 0.067 | 146 | 172 |
| $KBe_2BO_5H_2$[a)] | Cal. | *R32* | 6.63 | $d_{16}$=0.46 | 1.5675 | 1.5675 | 1.5105 | 0.057 | 188 | 250 |
| $Be_2BO_4H$[a)] | Cal. | *P1* | 7.62 | $d_{16}$=0.55 | 1.5816 | 1.5661 | 1.4964 | 0.085 | 163 | 183 |
| $Be_2BO_4H$ | Exp. | *Pbca* | --- | --- | 1.554-1.560 | 1.587-1.591 | 1.628-1.631 | 0.074 | --- | --- |
| | Cal. | *Pbca* | 7.77 | 0 | 1.5723 | 1.6002 | 1.6339 | 0.061 | 160 | N/A |
| $Be_2BO_5H_3$ | Exp. | | --- | --- | 1.580-1.582 | 1.580-1.582 | 1.485-1.493 | 0.095 | --- | --- |
| 1T-Phase | Cal. | *P3* | 8.20 | $d_{16}$=0.52 | 1.5683 | 1.5683 | 1.4816 | 0.087 | 152 | 175 |
| 2T-Phase | Cal. | *P3c1* | 7.67 | $d_{33}$=0.14 | 1.5986 | 1.5986 | 1.5064 | 0.092 | 162 | 185 |
| 2H-Phase | Cal. | $P6_3$ | 7.69 | $d_{33}$=0.25 | 1.5988 | 1.5988 | 1.5102 | 0.089 | 162 | 190 |

[a)] The compounds are theoretically designed.